\documentclass[fleqn,12pt,twoside]{article}
\usepackage{espcrc1}

\usepackage{graphicx}
\usepackage[figuresright]{rotating}

\newcommand{\AmS}{{\protect\the\textfont2
  A\kern-.1667em\lower.5ex\hbox{M}\kern-.125emS}}

\title{High $p_T$ $\pi^0$ Production and Angular Correlations in 158 $A$GeV
p+A and Pb+Pb Collisions}

\author{S. Bathe\address[MCSD]{University of M\"unster, 48149 M\"unster, Germany} for the WA98-Collaboration}%
\begin{document}

% typeset front matter
\maketitle

\begin{abstract}
Recent results of the WA98 experiment with p and Pb induced reactions
at 158 $A$GeV are presented.
Azimuthal $\gamma$-$\gamma$ correlations at high $p_{T}$ were studied
in search for a signal of jet-like structures.
A clear indication for back-to-back correlations can be seen in p+A
with a strong dependence on the $p_{T}$ of the photons and the size
of the
system. In Pb+Pb collisions in plane elliptic flow has been observed.
Results on transverse mass spectra of neutral pions measured at
central rapidity
are presented for impact parameter selected Pb+Pb collisions.
In going from peripheral to medium central collisions there is a
nuclear enhancement increasing with transverse mass similar to the
Cronin effect, while for
very central collisions this enhancement appears to be weaker than
expected.
\end{abstract}

\section{Introduction}
%The observation of a new phase of strongly interacting matter, the
%quark gluon plasma (QGP), is one of the most important goals of
%current
%nuclear physics research.
%The CERN experiment
%WA98~\cite{wa98proposal,wa98photonslong} searches
%for signatures of a phase transition to this new state
%of matter. 
%The experiment consists
The CERN experiment WA98~\cite{wa98proposal,wa98photonslong} consisted
of large acceptance photon and hadron spectrometers together with
several
other large acceptance devices which allow to measure various global
variables on an event-by-event basis.
The results presented here were obtained from an analysis of the
data taken with p and Pb beams in 1995 and 1996 at 158 $A$GeV.
The Pb-induced reactions
have been subdivided into samples of different centrality using the
transverse
energy $E_{T}$ measured in the MIRAC calorimeter.
Photons are measured with the WA98 lead-glass photon detector,
LEDA, which consisted of 10,080 individual modules with
photomultiplier readout. The detector was located at a distance of
21.5~m from the target and covered the pseudorapidity interval
$2.2 < \eta < 2.9$.
Details about the photon measurement can be found
in~\cite{wa98photonslong}.
%
%In this paper we will concentrate on particle production at high $p_T$ 
%where we discuss azimuthal $\gamma$-$\gamma$
%correlations and results on transverse mass spectra of neutral pions
%for different centralities.
\begin{figure}[ht]
\begin{center}
\includegraphics[scale=.8]{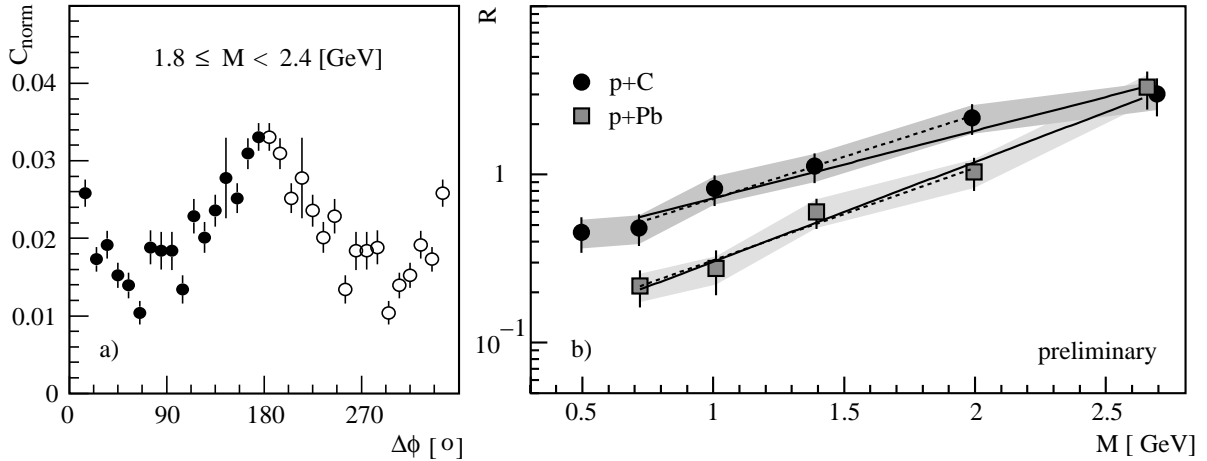}
\end{center}
\vspace{-1.5 cm}
\caption{a) Exemplary $C(\Delta \phi)$ for p+C reactions. 
b) Ratio $R$ of correlated to uncorrelated pairs
at $\Delta \phi=180^\circ$ as function of the pseudo-mass M for p+C and p+Pb.}
\label{fig1}
\end{figure}
\section{Azimuthal $\gamma$-$\gamma$-Correlations}
\begin{figure}[t]
\includegraphics[scale=.85]{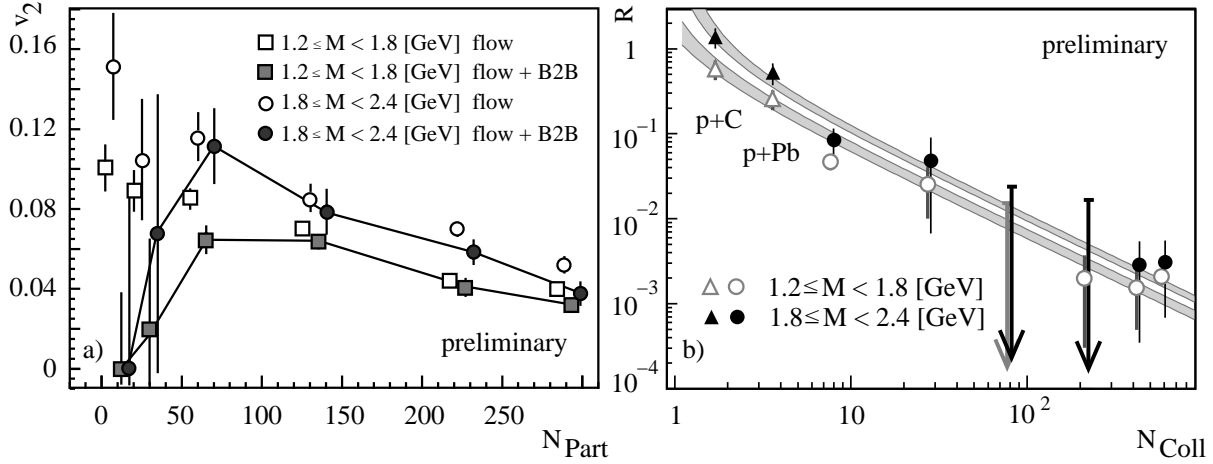}
\vspace{-1.5 cm}
\caption{a) $v_2$ in Pb+Pb as function of the number of 
participants with and without consideration of back-to-back effects.
b) Strength of back-to-back correlations in p+A and Pb+Pb reactions.}
\label{fig2}
\end{figure}

We attempt to use azimuthal correlations of
photons, which mainly originate from decays of neutral hadrons,
$\pi^{0}$ or $\eta$, to gain information on the relative fraction of
produced particles which still carry memory of the primary production
process. Particles produced in a primary two-body collision of incoming
nucleons must be correlated in transverse momentum due to
local momentum conservation. Dominantly, especially if
their individual transverse momenta are large, the particles will be
anti-correlated in azimuthal angle. This is of course reminiscent of
jet structure in the particle emission in high-energy physics.
In fact, there is expected to be a smooth transition of
this purely kinematical effect to mini-jet or jet production. At high enough
transverse momenta, such an analysis should allow to study minijet or
jet production and eventually allow to investigate effects such as
jet quenching. % which has been discussed as a promising signal for QGP
%formation. 
If the particles undergo secondary
and further rescatterings, as in an equilibrated system, the memory of
the correlation discussed above will be lost.
Of course, anti-correlation of particle pairs will also be generated
by collective effects like hydrodynamical elliptic flow. One therefore
has to check how much the measured effects might be altered by flow
phenomena.
The $\gamma$-pairs are characterized by the difference of
their azimuthal angle $\Delta \phi$ and the so-called pseudo-mass
$M=p_{T1}+p_{T2}$.
The correlation function C$(\Delta \phi)$ is introduced:
\begin{displaymath}
C(\Delta \phi) =
\left.
\frac{d^2N}{d\varphi_1d\varphi_2}\right|_{p_{T1}+p_{T2}}
\left/
\left(\frac{dN}{d\varphi_1}\frac{dN}{d\varphi_2}\right)\right|_{p_{T1}+p_{T2}}
\end{displaymath}
The combinatorial background of uncorrelated pairs is obtained by
event mixing
taking into account different centrality and multiplicity classes.
A clear correlation around $\Delta \phi=180^\circ$ can be
seen for p+C (Fig 1a) and p+Pb reactions.
%The same strong correlation around $\Delta \phi=180^\circ$ and the
%increase of the strength of the correlation with increasing
%pseudo-mass can be seen in
%HIJING 1.36~\cite{hijing} and VENUS 4.12~\cite{SIM-WERN-93-A} as well. This is
%noteworthy as VENUS doesn't contain explicit hard scattering.
It is assumed that the correlation originates from the direct
production of
$\pi^0$'s in single binary collisions and that its strength therefore
contains information about the ratio of directly produced to
rescattered
particles.
The correlation function for the p+A data can be described by a
Gaussian distribution, with the ratio $R$ of correlated to uncorrelated pairs
at $\Delta \phi=180^\circ$ increasing exponentially with
the pseudo-mass M as shown in Figure 1b.
Peripheral Pb+Pb data show a similar behaviour.
Semi-peripheral and semi-central Pb+Pb data show a correlation function
that is dominated by elliptic flow. The flow effects are evaluated
by means of a Fourier expansion \cite{Posk98,Oll92}.
\begin{eqnarray}
\frac{1}{N} \frac{dN}{d(\Delta\phi)}
= 1 + 2v^2_1{\rm cos}\Delta\phi + 2v^2_2{\rm cos}2\Delta\phi,
\end{eqnarray}
The Fourier coefficient $v_1$ quantifies
the directed flow, whereas $v_2$ quantifies the elliptic flow.
The Fourier coefficients $v_n (n = 1,2)$
can be extracted from the correlation function without determination
of the reaction plane, and hence no event plane resolution correction 
has to be applied.
Figure \ref{fig2}a shows $v_2$ for different cut-offs on the 
pseudo-mass as a function of different centralities in Pb+Pb with and without 
consideration of back-to-back effects.
%$v_2$ is of similar magnitude as results from NA49 using charged
%pions~\cite{na49}.
%A comparison with the event plane determined at target rapidity
%shows that the elliptic flow
%deduced from the photon pairs is oriented in the event plane~\cite{hubert}.
%Here the event plane is determined by using all fragments in
%the Plastic Ball detector.
Especially in the more peripheral samples, elliptic flow does not
describe the correlation completely. To compare the strength of an
additional Gaussian-like correlation for different systems 
the ratio $R$ of correlated to uncorrelated pairs
at $\Delta \phi=180^\circ$ is shown in Figure  \ref{fig2}b as a function
of the number of binary collisions. 
A parametrisation of the p+A data points with 
$R(N_{Coll})=1/(aN_{Coll}-1)$ is extrapolated to the Pb+Pb data, which 
represents scaling 
of the number of correlated pairs with $N_{Coll}$. No deviation 
from this scaling is observed.
\section{Centrality Dependence of $\pi^0$-Production}
Already from the experimentally determined shape of transverse mass
spectra
of hadrons it is evident that heavy ion reactions are
not merely a superposition of nucleon-nucleon collisions
\cite{wa80:pi0:98}.
In p+A collisions the flattening of the transverse mass spectra
compared to p+p (Cronin effect \cite{cronin}) has been
attributed to initial state multiple scattering of partons
\cite{LEV83}.
In the analysis of central reactions of Pb+Pb at
158 $A$GeV, however, it is
seen that both predictions of perturbative QCD
\cite{pqcd} and hydrodynamical parameterizations
%\cite{wa98:pi0:98,wang:1998:qcd} and hydrodynamical parameterizations
\cite{wa98:hydro:99} can describe the measured neutral pion spectra
reasonably well.
The understanding of the relative contributions of the various soft
and
hard processes in particle production is especially important in view
of the recent interest in the energy loss of partons in dense matter
\cite{jetq1,jetq2},
generally referred to as \emph{jet quenching}, as a possible probe
for the quark gluon plasma. Since one of the suggested experimental
hints
of jet quenching is the suppression of particle production at high
transverse momenta, it is important to understand other possible
nuclear modifications of particle production in detail.

%\begin{figure}[ht]
%\includegraphics[scale=.8]{alltogether5-a.eps}
%\vspace{-1.0 cm}
%\caption{Remember}
%\label{fig3}
%\end{figure}
More information in this respect
may be gathered from the variation of the particle spectra for
different reaction systems or different centralities~\cite{pi0new}.
Here we study the variations in absolute multiplicities.
Especially at high transverse momentum one na{\"{\i}}vely expects an
increase of the multiplicity proportional to the number of
collisions due to the importance of hard scattering.
In fact, it was already observed in p+A collisions
at beam energies of $ 200 - 400 \, \mathrm{GeV}$ \cite{cronin} that
the increase in cross section at high transverse momenta is even
stronger than the increase in the target mass.
The ratios of the measured pion multiplicity distributions for two
different samples (labeled X and Y)
normalized to the number of collisions
\begin{equation}
          \label{eqn:nuclmod1}
          R_{XY}(m_{T}) \equiv
	\frac{{\left( E\frac{d^3N}{dp^3} (m_{T}) / N_{coll} \right)} _{X}}
	{{\left( E\frac{d^3N}{dp^3} (m_{T}) / N_{coll} \right) }_{Y}}
\end{equation}
is introduced.

The ratio of peripheral Pb+Pb collisions to p+p increases
strongly with increasing transverse mass -- this is in line with the
Cronin effect discussed above. A similar trend is observed when going
from peripheral to medium-central data.
In addition, the pion production is seen to increase roughly
proportional to
the number of collisions
even at low transverse mass. Going from medium central to central
the trend is reversed: the ratio decreases with increasing transverse
mass and the
pion multiplicities increase more weakly than the number of
collisions.
The ratio of very central to central collisions shows an indication
of
a similar effect although not very significant.

Neither results of HIJING\cite{hijing} 
calculations nor a more refined pQCD calculation
\cite{wang:pqcd2} can reproduce the experimental data. Here the ratios are 
all $\ge 1$ and thus do not explain the centrality dependence observed. 


\begin{thebibliography}{9}
\bibitem{wa98proposal}WA98 Collaboration,
\newblock {\em Proposal for a large acceptance hadron and photon
spectrometer},1991,
\newblock Preprint CERN/SPSLC 91-17, SPSLC/P260;
%
%\bibitem{Peitzmann:1996:qm96}
WA98 Collaboration, M.M.~Aggarwal et~al., Nucl. Phys. {\bf A 610}, 200c
(1996).

\bibitem{wa98photonslong}
WA98 Collaboration, M.M.~Aggarwal et al,  nucl-ex/0006007,
submitted to Phys. Rev. C.

%%%%%%%%%%%%%%%%%%%%%%%%%%%%%%%%%%%%%%%%%%%%%%%%%%%%%%%%%%%%%%%%%%%%%%%%%%%%

%\bibitem{SIM-WERN-93-A}                        % VENUS 4.12
%      K.~Werner, Phys.~Rep.~{\bf 232}, 87 (1993).

\bibitem{Posk98}
      A.~M.~Poskanzer ans S.~A.~Voloshin, Phys. Rev. {\bf C58}, 1671 (1998). 

\bibitem{Oll92}
      J.~Y.~Ollitrault, Phys. Rev. {\bf D 46}, 229 (1992). 

%\bibitem{na49} H. Appelshaeuser et al., preprint nucl-ex/9711001.

%\bibitem{hubert} WA98 Collaboration, H. Schlagheck et~al., Nucl. Phys. {\bf A 661}, 337 (1999).
%%%%%%%%%%%%%%%%%%%%%%%%%%%%%%%%%%%%%%%%%%%%%%%%%%%%%%%%%%%%%%%%%%%%%%%%%%%%


\bibitem{wa80:pi0:98} WA80 Collaboration, R.~Albrecht et al.,
Eur. Phys. J. C {\bf 5}, 255--267 (1998).

\bibitem{cronin}
D.~Antreasyan, et~al., Phys. Rev. D {\bf 19}, 764 (1979).

\bibitem{LEV83}
M.~Lev and B.~Petersson, Z. Phys. {\bf C21} (1983) 155--161;
%\bibitem{krzywicki79}
A.~Krzywicki et al., Phys. Lett. {\bf B85}
    (1979) 407--416.
%\bibitem{wa98:pi0:98} WA98 Collaboration, M.~M.~Aggarwal et al.,
%Phys. Rev. Lett. {\bf 81} (1998) 4087; Phys. Rev. Lett. {\bf 84}
%(2000)
%578 (erratum).
%
%\bibitem{wang:1998:qcd} X.-N.~Wang, Phys. Rev. Lett. {\bf 81} (1998)
%2655-2658.

\bibitem{pqcd} WA98 Collaboration, M.~M.~Aggarwal et al.,
Phys. Rev. Lett. {\bf 81} (1998) 4087; Phys. Rev. Lett. {\bf 84}
(2000) 578 (erratum); X.-N.~Wang, Phys. Rev. Lett. {\bf 81} (1998)
2655.

\bibitem{wa98:hydro:99} WA98 Collaboration, M.~M.~Aggarwal et al.,
Phys. Rev. Lett {\bf 83} (1999) 926.


\bibitem{jetq1} X.-N.~Wang and M.~Gyulassy,
Phys. Rev. Lett. {\bf 68} (1992) 1480; X.-N.~Wang, Phys. Rev. C {\bf
58}
(1998) 2321.

\bibitem{jetq2} R.~Baier, D.~Schiff, and B.~G.~Zakharov
Ann.~Rev.~Nucl.~Part.~Sci. {\bf 50} (2000) 37-69 and references therein.

\bibitem{pi0new} M.~M.~Aggarwal et al.,
Eur.Phys.J. {\bf C 23} (2002) 225-236.

% \bibitem{wa98:scaling} WA98 Collaboration, M.M.~Aggarwal et al.,
% Eur. Phys. J. C {\bf 18} (2001),651--663.

%\bibitem{gyulassy98} M.~Gyulassy, and P.~Levai, Phys. Lett. B {\bf
%442} (1998) 1.

\bibitem{hijing}X.-N. Wang and M. Gyulassy, Phys. Rev. {\bf D 44},
           3501 (1991); Comp. Phys. Comm. {\bf 83}, 307 (1994).

\bibitem{wang:pqcd2} E.~Wang, and X.-N.~Wang, preprint
nucl-th/0104031.
\end{thebibliography}
\end{document}